# $Li_{14}Mn_2S_9$ and $Li_{10}Si_2S_9$ as a pair of all-electrochem-active electrode and solid-state electrolyte with chemical compatibility and low interface resistance

Qifan Yang[1,2], Jing Xu[1,3], Xiao Fu[1,2], Jingchen Lian[1,2], Liqi Wang[1,3], Xuhe Gong[1,4], Zibin Wang[1,3], Ruijuan Xiao[1,2,3,]* and Hong Li[1,2,3,]*

[1] Beijing National Laboratory for Condensed Matter Physics, Institute of Physics, Chinese Academy of Sciences, Beijing 100190, China

[2] Center of Materials Science and Optoelectronics Engineering, University of Chinese Academy of Sciences, Beijing 100049, China

[3] School of Physical Sciences, University of Chinese Academy of Sciences, Beijing 100049, China

[4] School of Materials Science and Engineering, Key Laboratory of Aerospace Materials and Performance (Ministry of Education), Beihang University, Beijing 100191, China

E-mail: rjxiao@iphy.ac.cn, hli@iphy.ac.cn

## Abstract

In solid-state batteries (SSBs), improving the physical contact at the electrode-electrolyte interface is essential for achieving better performance and durability. On the one hand, it is necessary to look for solid-state electrolytes (SSEs) with high ionic conductivity and no reaction with the electrode, on the other hand, to design the all-electrochem-active (AEA) electrodes that contain no SSEs and other non-active substances. In this work, we proposed a pair of AEA-electrode and SSE with the same structural framework and excellent interface compatibility, $Li_{14}Mn_2S_9$ and $Li_{10}Si_2S_9$, and confirmed the feasibility by ab-initio molecular dynamics (AIMD) simulations and machine learning interatomic potential based molecular dynamics (MLIP-based MD) simulations, providing a new approach to promote interfacial stability in SSBs.

# 1. Introduction

Compared with lithium-ion batteries (LIBs) that utilize liquid electrolyte, solid-state batteries (SSBs) with solid-state electrolytes (SSEs) are gaining wide attention due to their excellent performance[1]-[2]. As important materials in SSBs, the discovery and design of SSEs are of paramount importance[3]. Some SSEs with ionic conductivity that can reach the level of liquid electrolytes have been discovered, such as Li argyodites[4],[5], NASICON-type $LiM_2(PO4)_3$ (M = Ge, Ti, Sn, Hf, Zr)[6], $Li_{10}GeP_2S_{12}$ (LGPS)[7],[8], and garnet-type $Li_xLa_3M_2O_{12}$ ($5 \leq x \leq 7$, M = Nb, Ta, Sb, Zr, Sn)[9],[10]. However, numerous SSEs are impractical due to the interface problems caused by incompatibility between electrodes and electrolytes. Specifically, the primary reasons triggering the incompatibility between electrodes and SSEs including the chemical reactions and the physical contact failure between them[11]. On the one hand, most SSEs tend to easily react continuously and destructively with electrodes and form solid-electrolyte interphase (SEI) or cathode-electrolyte interphase (CEI) constantly[12]. On the other hand, during the cycling process of the battery, the volume of electrode changes a lot, and due to the flowability of liquid electrolytes, they are able to preserve intimate contact. However, for SSEs, the volume changes of the electrode create the space between the electrode and SSE, leading to the significant rise in polarization and interfacial resistance in batteries[13]. To address these issues, firstly, to improve the chemical compatibility, the solution is that choosing electrode materials that are inert to SSEs within their electrochemical window[14], [15]. Secondly, in order to mitigate the physical contact failure and reduce the interfacial resistance, researchers have made many attempts including expanding the contact area to improve mobile ion flux[16],[17], adding liquid phases to lubricate the interface[13], constructing interfaces with close atomic contact by epitaxial growth[18] and so on. Creatively, Li[19] et al. introduced the concept of all-electrochem-active (AEA) electrode with both excellent electronic and ionic conducting performance aiming to eliminate the unstable interfaces inside the electrode without adding SSEs and other additives. Wang[20] et al. presented a single cell with only $Li_3TiCl_6$ material as both electrode and SSE, thus avoiding the impact brought by the interfaces between electrodes and SSEs. Based on these previous explorations, we further speculated that identifying and designing pairs of electrodes and SSEs with the same structural framework possessing great $Li^+$ ion transport properties that exhibit both chemical compatibility and low interfacial resistance can also improve the interfacial compatibility, because the pairs can minimize interfaces within the AEA-electrode and increase the physical and chemical stability at the interface between AEA-electrode and SSE without adding SSE into the electrode which might reduce the capacity of SSBs.

In this work, a pair of AEA-electrode and SSE with the same structural framework is chosen to validate the strategy for designing interfacial compatibility. $Li_{14}Mn_2S_9$ (LMS), acting as an AEA-electrode, was obtained by high-throughput screening based on the structural characteristic involving isolated anions that we discovered recently[21]. LMS is thermodynamically stable and exhibiting Li$^+$ ion transport behavior, and its substantial Li content, the presence of transition metal element Mn as well as the great Li$^+$ ion conducting property, suggest a strong potential for the application as an AEA-electrode material. Additionally, through element substitution, we predicted a new SSE material Li$_{10}$Si$_2$S$_9$ (LSS) sharing the same structural framework with LMS exhibiting high stability and ionic conductivity. The same structural framework implies that they are likely to form a structurally simple and stable interfacial atomic layer at the contact area.To prove this probability, we conducted simulations on LMS and LSS structures through ab-initio molecular dynamics (AIMD), which is frequently used to simulate structural evolution and ionic transport of crystalline systems[22]. However, to simulate the interfaces of electrode and SSE, AIMD method encounters some troubles. First, the interface model is typically the huge model with over 1,000 atoms, yet AIMD method is usually used to conduct simulations in systems with limited total number of atoms (generally fewer than 200 atoms) due to the great computational cost. Second, the interface structure possibly evolves over time due to interface reactions, causing a timescale that exceeds the range of AIMD simulates. Moreover, the ion transport mechanisms of the interface at high- and low-temperatures may differ, thus the room-temperature properties, e.g. the ionic conductivity, can't be extrapolated from high-temperature simulations. Therefore, to perform simulations on interface models with longer time, the machine learning interatomic potential (MLIP) is employed in this study through the recently presented software Neural Equivariant Interatomic Potentials (NequIP)[23], which is a high-performance method to train MLIP models and is promising for structural and ion conducting simulations on interfaces of materials.

We proved the viability of LMS as an AEA-electrode and LSS as a SSE by the analysis of their structural and ion conducting properties and the delithiated products of LMS performed via density functional theory (DFT) calculations and AIMD simulations. In addition, we used NequIP to train a MLIP model and conducted molecular dynamics (MD) simulations of the interfaces between LMS and LSS to investigate the structure and ion transport mechanisms at the interfaces. Simultaneously, we provide the electronic density of states (DOS) of surfaces and interfaces for LMS and LSS. The results illustrate that LMS is a qualified AEA-cathode and LSS exhibits the extraordinary properties of SSE. Moreover, the LMS-LSS pair shows a high degree of chemical compatibility, and their interfaces possess high ionic conductivity. Thus, we have adopted an effective method by introducing a compatible pair of AEA-electrode and SSE, LMS-LSS, to reduce the chemical reactions and interfacial resistance between electrodes and SSEs in batteries which can significantly affect battery performance.

## 2. Materials and Methods

**2.1 Structure acquisition:** Structure of $Li_{14}Mn_2S_9$ (LMS) was obtained from Materials Project database (mp-756198) by high-throughput screening process characterized by isolated anions[22]. Structure of $Li_{10}Si_2S_9$ (LSS) was derived by element substitution of $Li_{14}Mn_2S_9$. Firstly, the Mn atoms in LMS were all replaced by Si atoms. Then, in order to produce 4 $Li^+$ vacancies in the cell, using enumeration method and taking the Wyckoff positions of atoms as well as Ewald energy into account, 50 distinct structures were generated and relaxed by density functional theory (DFT) calculations. Ultimately, the structure with the lowest calculated total energy was selected as the configuration for LSS. Similarly, a series of the structures of delithiated products were obtained by enumeration, structural relaxation, and choosing the configuration with the lowest energy. The relevant information on the construction of surface and interface structure models is displayed in the Note S1. Visualization of these structures was accomplished using the VESTA software package[24]. The enumeration process of structures was conducted using the pymatgen and the pymatgen.transformations.advanced_transformations Python packages.

**2.2 Density functional theory computation:** We conducted all relaxation calculations and DOS calculations within unit-cells (refer to Table S3 for the atom number of each cell) employing Vienna Ab initio Simulation Package (VASP)[25], grounded in density functional theory (DFT). We used Perdew–Burke–Ernzerhof (PBE)[26] generalized gradient approximation (GGA) in conjunction with projector-augmented-wave (PAW) approach. However, the intercalation voltages were calculated by Heyd-Scuseria-Ernzerhof (HSE06) hybrid functional[27], [28], the results of which was proved to be closer to experimental values compared with GGA[29]. What's more, the wavefunction and density cutoffs are 520eV and 780eV. During the optimization, both ions and cells were relaxed, with convergence criteria for energy and force set at $10^{-5}$ eV and 0.01 eV/Å, respectively. The thermodynamic stability was evaluated using $E_{hull}$, derived from phase diagrams within the Materials Project (MP) database[30], [31].

**2.3 Ab initio molecular dynamics simulation:** AIMD simulations were conducted for supercells of both LMS and LSS to study their ion transport properties (refer to Table S3 for the atom number of each cell). These simulations employed supercells with lattice parameters greater than or approximately 10 Å, and utilized nonspin-polarized DFT calculations for LSS and its surfaces, as well as spin-polarized DFT for LMS, its surfaces and all interfaces, with a Γ-centered k-point mesh. We performed the AIMD calculations with a Nose thermostat[32] for 70,000 steps at 700K, 800K, 900K and 1000K, and the time step is 1 fs. The initial 10 ps were allocated for structural equilibration, while the subsequent 60 ps were utilized to statistically determine the ion migration properties. All analysis were executed using the pymatgen and the pymatgen-diffusion Python packages.[33], [34]

**2.4 MLIP-based molecular dynamics simulation:** MLIP models was trained using NequIP[20]

and MD simulations of the interfaces of LMS and LSS were performed utilizing Large-scale Atomic/Molecular Massively Parallel Simulator (LAMMPS)[35] in supercells (refer to Table S3 for the atom number of each cell). The information about MLIP model training is included in Note S2. These MD simulations were conducted under constant volume and temperature (NVT) conditions. The MD simulations were conducted for 1,000,000 steps at 330K or 400K, employing the time step of 1 fs. For MD simulations, the 0 ps to 1,000 ps was utilized for subsequent analysis. All analysis were conducted by pymatgen and the pymatgen-diffusion Python packages. The open visualization tool (OVITO)[36] was employed to visualize and analyze the results from LAMMPS.

## 3. Results and discussion

### 3.1 Structures, thermodynamic stability and bond valence (BV) analysis

Firstly, we obtained the structures of both LMS and LSS. LMS structure, shown in Table 1, was discovered by the high-throughput screening from Materials Project database according to the concept of isolated anions as we discussed in previous research[21]. As shown in Fig. 1a, the structure is composed of isolated S anions ($S_{iso}$) encircled solely by $Li^+$ ions, along with the surrounding $Li^+$ ions, and the $MnS_4$ tetrahedron. LSS structure is derived from the LMS structure by substituting Si atoms for Mn atoms, and subsequently removing a number of lithium atoms to maintain valence equilibrium, thus it keeps the same structural framework as LMS with $S_{iso}$, $Li^+$ ions and $SiS_4$ tetrahedron (Fig. 1b). However, compared with LMS, the substitution of Si elements and the rise in $Li^+$ vacancies result in a decrease in the overall volume of the unit-cell of LSS (Table 1) and changes in lattice parameters (Table S1). The substitution process is detailed in methods section. Secondly, the thermodynamic stability of both materials is assessed using the convex hull energy ($E_{hull}$) and phase diagram analysis. The Materials Project database offers the $E_{hull}$ value of the LMS which is only < 1 meV/atom (We also calculated the $E_{hull}$ value of the LMS which is 0 meV/atom, indicating the consistency between our results and the Materials Project database calculated results.), ensuring the thermodynamic stability. For the LSS which is not included in Materials Project database, we conducted an analysis of $Li_2S$-S-Si ternary phase diagram as depicted in Fig. 1c and determined that its $E_{hull}$ value is 0 meV/atom, indicating its thermodynamic stability. Furthermore, based on our previous research, it has been concluded that the presence of $S_{iso}$ triggers the frustration phenomenon and thereby induces $Li^+$ cage transport surrounding $S_{iso}$ and enhances ionic conductivity, thus the migration energy barriers of $Li^+$ transport and isosurfaces used as $Li^+$ diffusion pathways of LMS and LSS predicted by the bond valence (BV)[37] method are calculated to roughly analyse their ionic transport properties. The migration energy barriers presented in Table 1 are all below 1 eV, indicating both materials may possess great ion conducting properties. Specifically, the migration energy barrier for LSS (0.55 eV) is lower than LMS (0.97 eV), which means that the ionic conductivity of LSS is likely to be higher than that of LMS. What's more, the $Li^+$ diffusion pathways for both materials are three-dimensional and exhibit cages around $S_{iso}$ as predicted (Fig. 1d and 1e).

Thus, through high-throughput screening and element substitution, we identified the structures of both LMS and LSS which share one structural framework with minor differences in structures. Also, both materials exhibit great thermodynamic stability, indicating a high likelihood of successful synthesis. According to the preliminary inference from the BV method, both of the compounds have favorable ion transport performance and a cage transport mechanism, with LSS demonstrating superior ion conducting property than LMS.

Table 1: Basic information of LMS and LSS.

| Structure | Origination | Unit-cell volume (Å$^3$) | Migration energy barrier calculated by BV method (eV) |
|---|---|---|---|
| Li$_{14}$Mn$_2$S$_9$ | mp-756198 | 424.49 | 0.97 |
| Li$_{10}$Si$_2$S$_9$ | Li$_{14}$Mn$_2$S$_9$ | 413.38 | 0.55 |

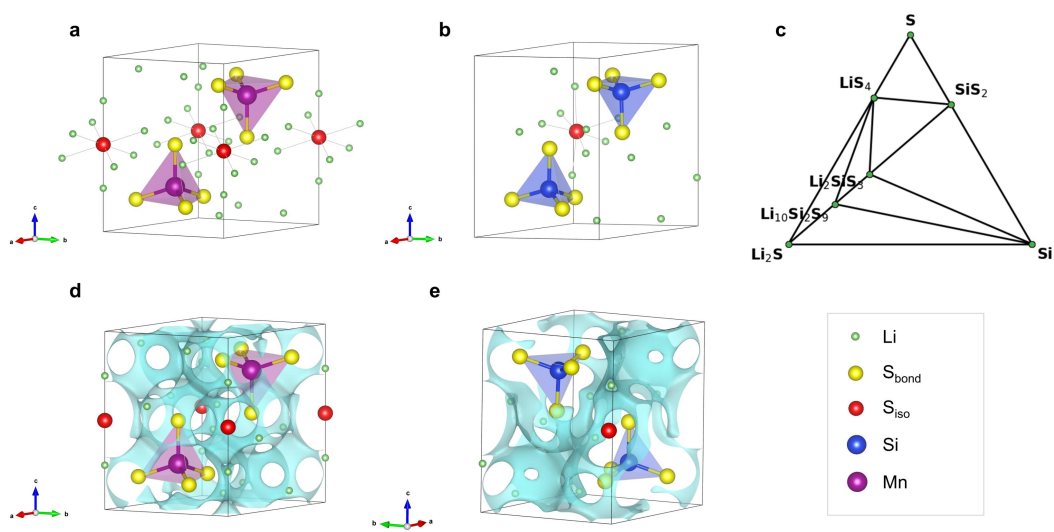

Fig. 1: The crystal structures and ionic migration properties of LMS and LSS. **a** Structure of LMS and **b** Structure of LSS relaxed by DFT calculations. **c** Li$_2$S-S-Si ternary phase diagram. **d** The Li$^+$ diffusion pathway in LMS and **e** The Li$^+$ diffusion pathway in LSS evaluated by BV method with the isosurface of E = 0.97 eV and 0.55 eV, respectively. The red balls refer to the isolated S (S$_{iso}$) atom, and the yellow balls refer to the bonded S (S$_{bond}$) atoms. The green, purple and blue balls refer to the Li$^+$ ions, Mn atoms and Si atoms.

### 3.2 Li$^+$ ion transport performances

To further analyze the Li$^+$ ion transport performance of LMS and LSS more accurately, AIMD simulations at 700K, 800K, 900K and 1000K were conducted. The resulting mean squared displacements (MSDs), diffusion pathways as well as Arrhenius curves derived from AIMD simulations for both materials are illustrated in Fig. 2 and the Li$^+$ ion diffusivity, ionic conductivity at 300K, and the ionic migration barriers ($E_a$) derived from Arrhenius curves for LMS and LSS are presented in Table 2. First of all, as depicted in Fig. 2b, for LMS, Mn and S atoms remain unmoved at all temperature ranges. Regarding LSS (Fig. 2e), it can be observed that Si atoms retain their positions at different temperatures, and the bonded S ($S_{bond}$) atoms rotate around their connected Si atoms, which can be confirmed by the unchanged Si-S bond length shown in Fig. S1 and the rotating pathway of $S_{bond}$ atoms showed in Fig. 2g, indicating the stability of their structural framework and the precision of subsequent analysis.

Regarding $Li^+$ ion movement, we obtained the ion conductivities of LMS and LSS at room temperature by interpolating the results of AIMD simulations at various temperatures mentioned above. Notably, the ion conductivity of LSS reaches 2.01 mS/cm at 300K, meeting the requirements of SSEs. What's more, although LMS can also reach high ionic conductivity of 0.011 mS/cm at 300K which ensures its ability to become an AEA-electrode, the $Li^+$'s diffusivity and ion conductivity of LSS are two orders of magnitude higher than those of LMS, and LSS also has more obvious cage characteristics than LMS at the same temperature. Compared with LMS, a slight decrease in unit-cell parameters, an increase in the number of $Li^+$ vacancies and the rotation of $S_{bond}$ atoms appear in LSS. More $Li^+$ vacancies and the synergy between the rotation of $S_{bond}$ and the motion of $Li^+$ may be the reason of the better ionic conducting property of LSS. To explore the structural reasons for the $S_{bond}$ atoms' rotation in LSS, we examined the bonding characteristics of two structures. As illustrated in Fig. S2, within LMS structure, the local environments of all S atoms bonded with Mn are very similar, as their Mn-S bond lengths are all 2.38 Å and they are all 6-coordinated with Li. However, in LSS, a reduction in lithium content and changes in its arrangement result in a decrease and non-uniformity in the number of $Li^+$ ions surrounding $S_{bond}$ atoms, leading to two different chemical environments of $S_{bond}$, S1 and S2. The distances between them and Si have a difference of 0.04 Å, with Si-S1 bond being 2.14 Å and Si-S2 bond being 2.18 Å. In addition, S1 coordinates with 3 or 4 $Li^+$ ions, while S2 coordinates with 5 $Li^+$ ions. We also discovered that during the AIMD simulations for LSS structure, only S1 rotates while S2 remains fixed. This phenomenon could be attributed to the lower coordination numbers of S1 compared to S2 or S in LMS, resulting in weaker attraction forces and thus a greater probability for rotations.

As a result, LMS and LSS both have good ion transport properties, with conductivity of 0.011 mS/cm and 2.01 mS/cm at 300K, respectively. The ion transport performance of LSS can meet the criteria for SSEs, and the high ionic conductivity makes LMS a potential candidate for AEA-electrode. Due to the structural disparities between the two compounds, rotation of $S_{bond}$ atoms as well as more Li vacancies occur in LSS, facilitating the superior ionic conductivity.

Table 2: The Li$^+$'s diffusivity and ionic conductivity at 300K extrapolated through Arrhenius curves by AIMD simulations for LMS and LSS.

| Structure | Diffusivity (cm$^2$/s) | Ionic conductivity at 300K (mS/cm) | E$_a$ (eV) |
|---|---|---|---|
| Li$_{14}$Mn$_2$S$_9$ | 1.354 × 10$^{-10}$ | 0.011 | 0.46 |
| Li$_{10}$Si$_2$S$_9$ | 1.343 × 10$^{-8}$ | 2.01 | 0.31 |

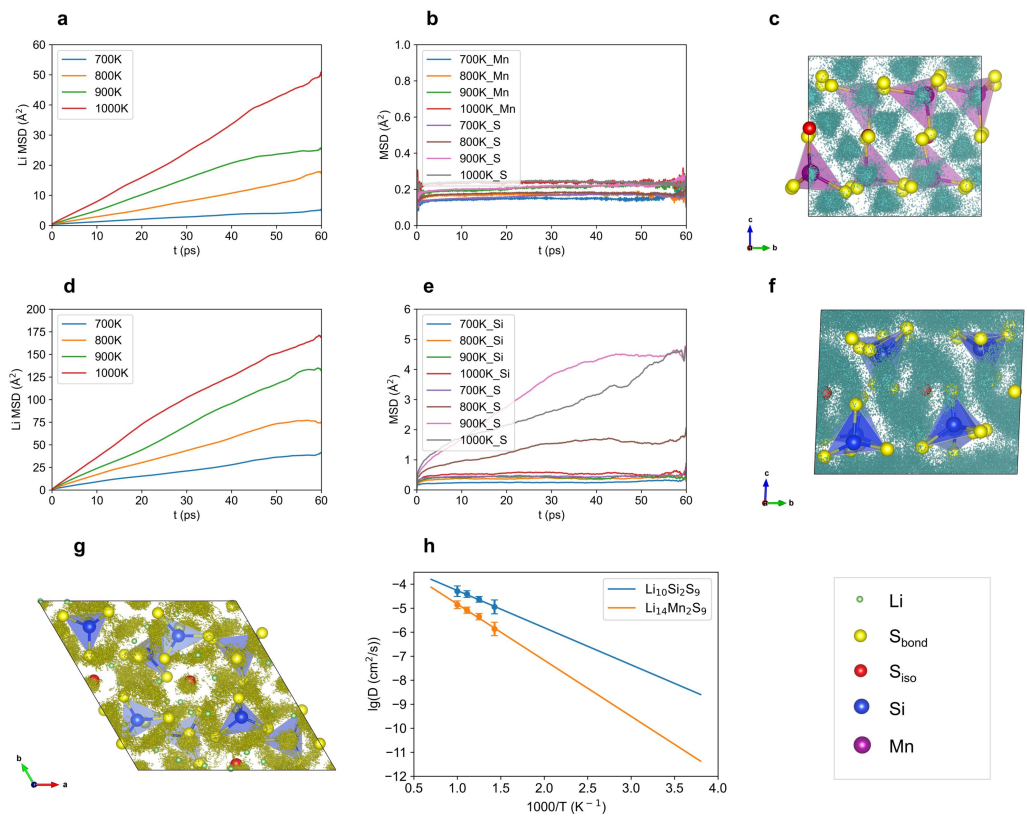

Fig. 2: The ionic transport properties of LMS and LSS calculated by AIMD. **a** Li$^+$'s MSDs for LMS at 700K, 800K, 900K and 1000K. **b** Mn and S atoms' MSDs for LMS at 700K, 800K, 900K and 1000K. **c** Li$^+$ diffusion pathway at 1000K in LMS obtained from AIMD simulations. **d** Li$^+$'s MSDs for LSS at 700K, 800K, 900K and 1000K. **e** Si and S atoms' MSDs for LSS at 700K, 800K, 900K and 1000K. **f** Li$^+$ diffusion pathway at 1000K in LSS obtained from AIMD simulations. **g** S atoms' diffusion pathway at 1000K in LSS obtained from AIMD simulations. **h** Arrhenius curves of LMS and LSS fitted with AIMD data. The red balls refer to the S$_{iso}$ atoms. The yellow balls refer to the S$_{bond}$ atoms. The green, purple and blue balls refer to the Li$^+$ ions, Mn atoms and Si atoms.

### 3.3 Delithiated products and intercalation voltage of LMS

Given that LMS material contains the transition metal Mn element which is commonly found in cathode materials, along with a high concentration of Li$^+$ ions as well as high ionic conductivity confirmed in above analysis, we proceed to further explore the potential of utilizing LMS as an AEA-cathode. Initially, we obtained the configurations of various delithiated products of LMS through introducing Li$^+$ vacancies, as described in the methods section. At the same time, we also calculated the E$_{hull}$ and corresponding intercalation voltages of each delithiated product, as presented in Table 3. The results indicate that as the quantity of lithium removal rises, the stability of the delithiated products as well as the intercalation voltage gradually decrease. When the lithium removal amount x is less than 5 atoms/cell, the structural framework of the delithiated products remains the same as LMS (Fig. S3), and the E$_{hull}$ value stays under 100 meV/atom, signifying possible stability during cycling. At this time, the intercalation voltage value ranges from 3.43 V to 3.14 V, which is relatively high and satisfies the requirements for use as a cathode. Additionally, in Fig. 3, we obtained the electronic density of states (DOS) for LMS and delithiated products, and the results indicate that even though the LMS exhibit insulating properties with the energy bandgap of 1.35 eV, none of its delithiated products exhibit an energy gap at the Fermi level, suggesting they are all conductors, which can conducting electrons during charging/discharging processes, fulfilling the criteria for electrode applications. What's more, it can be clearly seen that as the amount of lithium removal increases, the DOS of Mn at the Fermi level (or the peak closest to the Fermi level) decreases greatly and the DOS of S at the Fermi level decreases slightly, indicating that the valence of Mn element increases and Mn element mainly acts as the charge compensation.

Through above studies, we obtained the structures of delithiated products of LMS and assessed their stability, DOS, and theoretical intercalation voltage. The findings suggest that when the lithium removal amount is less than 5 atoms/cell, the delithiated product has commendable stability, electronic conductivity, and an appropriate theoretical voltage, which also proves the feasibility of LMS as an electrode.

Table 3: Delithiated products and their corresponding E$_{hull}$ and intercalation voltage evaluated by DFT calculations.

| Delithiated products | E$_{hull}$ (eV/atom) | Intercalation voltage (V)(by HSE06) |
| --- | --- | --- |
| Li$_{13}$Mn$_2$S$_9$ | 0.020 | 3.43 |
| Li$_{12}$Mn$_2$S$_9$ | 0.050 | 3.14 |
| Li$_{10}$Mn$_2$S$_9$ | 0.084 | 3.14 |

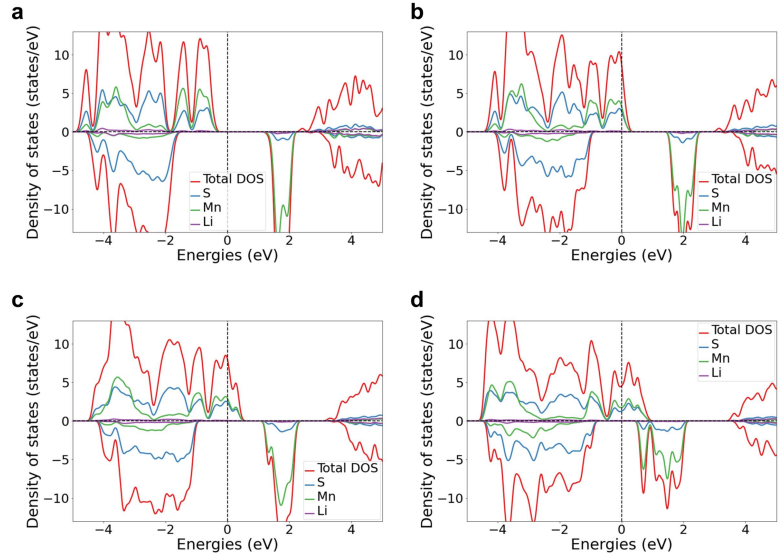

Fig. 3: The electronic density of states (DOS) for **a** $Li_{14}Mn_2S_9$, **b** $Li_{13}Mn_2S_9$, **c** $Li_{12}Mn_2S_9$, and **d** $Li_{10}Mn_2S_9$.

**3.4 The interfaces between LMS and LSS**

The analysis of the above results indicates that LMS is a promising candidate for AEA-electrode materials, and LSS could serve as SSEs. To investigate the stability and $Li^+$ ion transport at the interface between LMS and LSS, we further utilized DFT method and MLIP-based MD method with NequIP software to construct surface and interface models of both materials and perform comprehensive simulations on the interface models to analyze their characteristics. We chose the (100) and (001) surface of LMS and LSS (LMS-(100), LMS-(001), LSS-(100), LSS-(001)) to perform the DFT calculations, and for the interface models, besides the original (100) and (001) interfaces, LMS/LSS-(100) and LMS/LSS-(001) with overall composition of $Li_{12}MnSiS_9$, we also studied the delithiated interface structure for (100) interface, de-LMS/LSS-(100) model with overall composition of $Li_{11}MnSiS_9$, to investigate the change at interfaces caused by $Li^+$ extraction. The relevant information on the construction of surface and interface structure models is displayed in the Note S1. The relevant structural models are presented in Fig. S4 and S5.

Here, we simulated the interfaces between LMS and LSS using MLIP-based MD. The information about MLIP model training is included in Note S2. The MD simulations for LMS/LSS-(100) were conducted at low temperatures of 270K, 300K, 330K and 400K and for LMS/LSS-(001) were at 330K and 400K. Simultaneously, the de-LMS/LSS-(100) was simulated at 400K for comparison. Each interface structural model can be divided into three distinct phases: LMS phase, interfacial phase, and LSS phase, as Fig. S6 and S7 show. In order to characterize the migration characteristics of $Li^+$ ions in different phases, we define their types based on the initial region of each $Li^+$ ion and calculate their MSD over the simulation time. The $Li^+$ MSDs and diffusion pathways of each phase for the LMS/LSS-(100) and LMS/LSS-(001) at 330K and 400K, as well as the de-LMS/LSS-(100) at 400K during 1000 ps, are displayed in Fig. 4, Fig. 5, Fig. S6 and Fig. S7. We also displayed Li$^+$ MSDs at 260K, 300K, 330K, 370K and 400K for the LMS/LSS-(100) to observe the effect of temperature on Li$^+$ ion transport capability (Fig. S8). Firstly, it is noteworthy that in all simulations for interfaces the MSDs of Si, Mn, and S atoms remained near to zero as Fig. 4 and Fig. 5 show, indicating the remarkable stability of the structures, with no signs of collapse. Besides, the formation energy for one Mn/Si antisite at interface is 0.525 eV, 0.853 eV for LMS/LSS-(100) and LMS/LSS-(001), respectively (Table S2), indicating inter-diffusion phenomenon or chemical reactions is difficult to occur at the LMS/LSS interfaces, implying the chemical compatibility between LMS and LSS. However, the calculated formation energy for one Mn/Si antisite at de-LMS/LSS-(100) interface is -0.048 eV (Table S2), which means Mn-Si diffusion may occur in small quantities at the delithiated interface. Furthermore, it is found from Li$^+$ MSDs that in the LMS/LSS interface structures, Li$^+$ transport is contributed mostly by Li$^+$ transport in LSS phase and a few transport in the interface phase, while in LMS phase, Li$^+$ ions are undergoing only localized vibrations (Fig. 4a, 4c and 5). Fig. S6a, S6c and S7 also illustrate the phenomenon that Li$^+$ ions only transport in LSS phase and the interface phase but remain no diffusion in LMS phase from the perspective of Li$^+$ diffusion pathways. However, in de-LMS/LSS-(100), Li$^+$ ions' migrations in three phases are all efficient accompanied by the enhancement of Li$^+$ ions' transport in LMS phase (Fig. 4b), and the Li$^+$ transport capability in these phases can be ranked as follows: LSS phase > interface phase > LMS phase. From Fig. S7b, it can also be seen that Li$^+$ in the LMS phase no longer vibrates only at its original position, but achieves mutual transport within the LMS phase and the interface phase. Li$^+$ ions in the interface phase can also transport to the LMS and LSS phase, achieving the Li$^+$ ion transports in the entire structure. The calculated Li$^+$ ionic conductivity of interface structural models at 400K are 23.7 mS/cm for LMS/LSS-(100), 14.7 mS/cm for LMS/LSS-(001), and 54.5 mS/cm for de-LMS/LSS-(100), confirming the low interfacial resistance at 400K. Overall, the interfaces between LMS and LSS in original structures exhibit good Li$^+$ ion transport capabilities mostly contributed by LSS and the interface phase, and in the delithiated state, the Li$^+$ transport in LMS phase is triggered, forming the Li$^+$ movement in the entire structure, even at low temperatures. The simulations indicate that the two materials sharing the same structural framework allow for smoother connection and transportation of Li$^+$ at the interface.

From the analysis of the interfaces, we have determined that LMS and LSS exhibit excellent chemical compatibility without undergoing chemical reactions. And there is great ion transport ability at the interface, reducing high interfacial resistance. Therefore, LMS-LSS can serve as a well compatible AEA-electrode and solid-state electrolyte pair, and if successfully implemented in SSBs, they are promising candidates to mitigating the existing interfacial challenges.

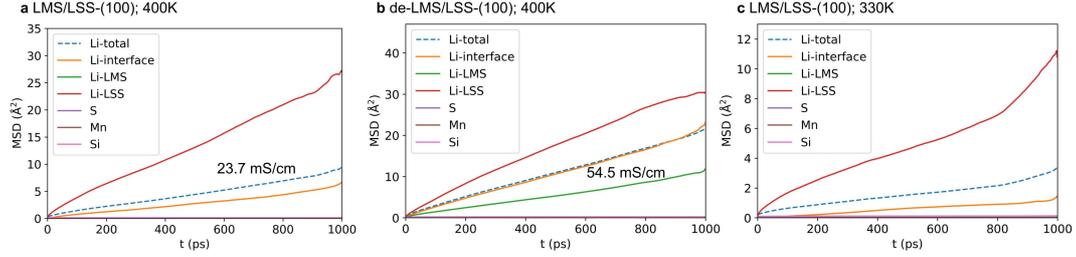

Fig. 4: The MSDs of LMS/LSS-(100) calculated by MLIP-based MD. **a** The MSDs of each element and Li$^+$ MSDs of the three phases for LMS/LSS-(100) at 400K. **b** The MSDs of each element and Li$^+$ MSDs of the three phases for de-LMS/LSS-(100) at 400K. **c** The MSDs of each element and Li$^+$ MSDs of the three phases for LMS/LSS-(100) at 330K.

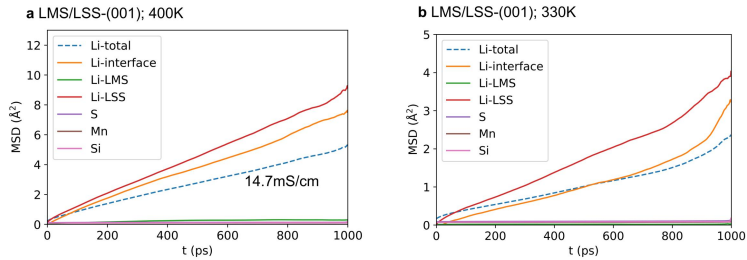

Fig. 5: The MSDs of LMS/LSS-(001) calculated by MLIP-based MD. **a** The MSDs of each element and Li$^+$ MSDs of the three phases for LMS/LSS-(001) at 400K. **b** The MSDs of each element and Li$^+$ MSDs of the three phases for LMS/LSS-(001) at 330K.

**3.5 The electronic density of states (DOS) of surfaces and interfaces for LMS and LSS**

To further examine the electronic structure of all structural models, the DOS and the partial electronic density of states (PDOS) were calculated and depicted in Fig. 6. Initially, as showed in Fig. 6a and 6b, both LMS and LSS exhibit insulating properties with energy bandgaps at the Fermi level, making LSS aligns with the characteristics of SSEs. Notably, due to the fact that the delithiated products of LMS are all conductors, its insulation property does not impede its efficacy as an electrode. Additionally, the DOS and PDOS of surfaces are similar to their respective bulk structures (Fig. 6c-f) as predicted. And the DOS of LMS/LSS-(100) and LMS/LSS-(001) (Fig. 6g and 6h) indicates that they possess semiconductor characteristics, yet the DOS of de-LMS/LSS-(100) with a rightward offset compared with LMS/LSS-(100) as showed in Fig. 6g and 6i, which are consistent with the DOS of bulk LMS and its delithiated product (Fig. 3a and 3b), shows conductor's performance of de-LMS/LSS-(100). Thus, from a DOS perspective, LMS fulfills the criteria for electrodes, while LSS satisfies the requirements for SSEs.

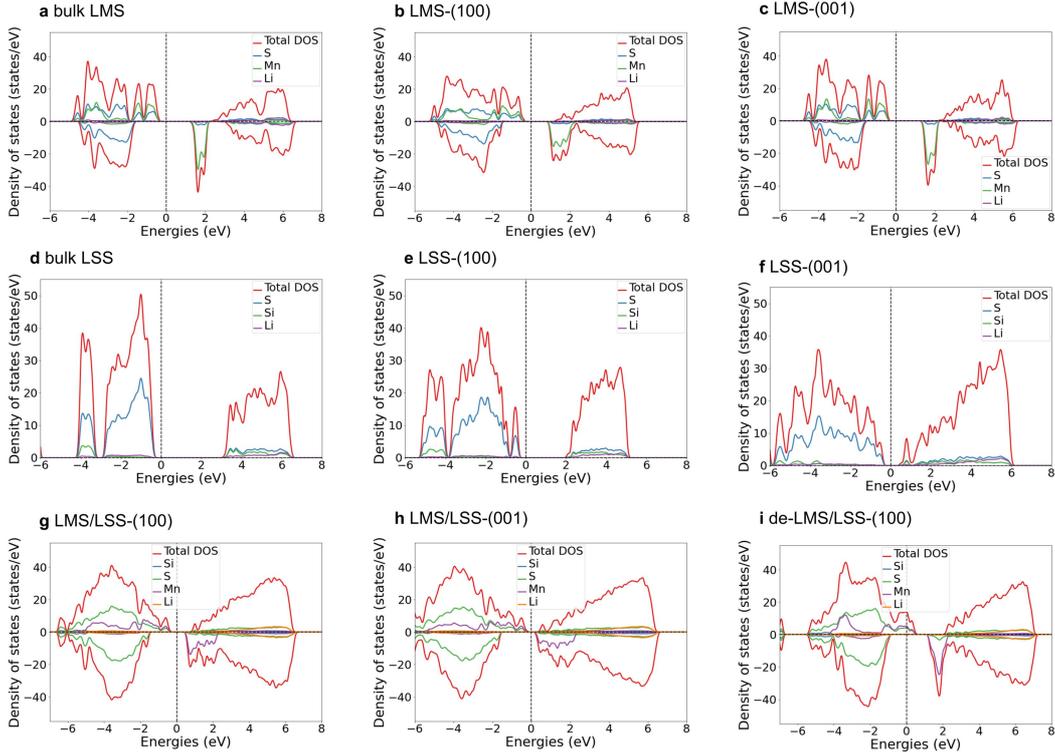

Fig. 6: The electronic density of states (DOS) of **a** bulk LMS, **b** LMS-(100) surface, **c** LMS-(001) surface, **d** bulk LSS, **e** LSS-(100) surface, **f** LSS-(001) surface, **g** LMS/LSS-(100) interface, **h** LMS/LSS-(001) interface, and **i** de-LMS/LSS-(100) interface.

## 4. Discussion

In conclusion, we propose a novel AEA-electrode and solid-state electrolyte pair, $Li_{14}Mn_2S_9$ (LMS) and $Li_{10}Si_2S_9$ (LSS), which shows promise in mitigating the current challenges of interface reactions and high interfacial resistance in solid-state batteries. Both LMS and LSS demonstrate thermodynamic stability. As an electrode, LMS yields stable and highly electronically conductive delithiated products, with appropriate intercalation voltage. As a SSE, LSS exhibits high room-temperature ionic conductivity of 2.01 mS/cm. Compared with LMS, additional Li vacancies and different chemical environments in LSS lead to the rotation of S atoms, and promote the transport of $Li^+$. Additionally, the interface between the two materials has excellent chemical compatibility and superior $Li^+$ ion transport capabilities, rendering them a potential candidate for compatible AEA-electrode and SSE pairs. The discovery and analysis of LMS and LSS provide a new and effective approach for tackling interface issues in solid-state batteries. Simultaneously, we also presents a way to screen electrodes and electrolytes with interface compatibility through this work. However, the LMS and LSS pair we selected still has the problem of not being able to fully utilized the lithium capacity in LMS, because even though LMS has 14 Li atoms in one unit-cell, when the lithium removal amount exceeds 6 atoms/cell, the framework structure collapses. Additionally, although there is no chemical reaction occurring at the interface of LMS/LSS, Mn-Si inter-diffusion may happen at the delithiated interface. To solve these problems, doping and modification methods can be used later to attempt to improve the structural stability of LMS as well as the chemical stability of the delithiated LMS/LSS interface during battery cycling, and looking for other AEA-electrode and SSE pairs are encouraged.

# Appendix A. Supporting information

Supplementary data associated with this article can be found in the online version at ****.

# Supporting Information for

# Li$_{14}$Mn$_2$S$_9$ and Li$_{10}$Si$_2$S$_9$ as a pair of all-electrochem-active electrode and solid-state electrolyte with chemical compatibility and low interface resistance

Qifan Yang[1, 2], Jing Xu[1, 3], Xiao Fu[1, 2], Jingchen Lian[1, 2], Liqi Wang[1, 3], Xuhe Gong[1, 4], Zibin Wang[1, 3], Ruijuan Xiao[1, 2, 3], * and Hong Li[1, 2, 3], *

[1] Beijing National Laboratory for Condensed Matter Physics, Institute of Physics, Chinese Academy of Sciences, Beijing 100190, China

[2] Center of Materials Science and Optoelectronics Engineering, University of Chinese Academy of Sciences, Beijing 100049, China

[3] School of Physical Sciences, University of Chinese Academy of Sciences, Beijing 100049, China

[4] School of Materials Science and Engineering, Key Laboratory of Aerospace Materials and Performance (Ministry of Education), Beihang University, Beijing 100191, China

E-mail: rjxiao@iphy.ac.cn, hli@iphy.ac.cn

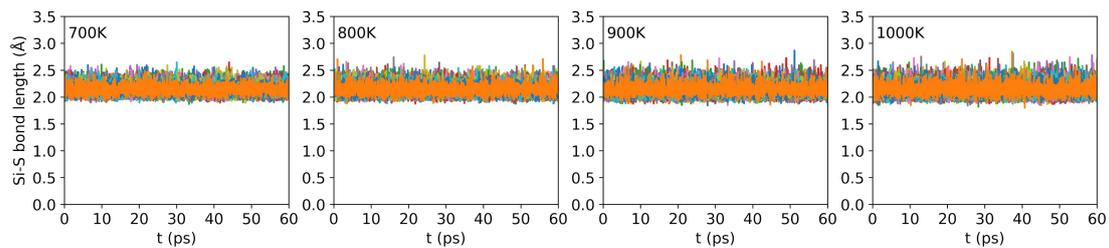

Fig. S1. The Si-S bond length of $Li_{10}Si_2S_9$ (LSS) in AIMD simulations at different temperatures.

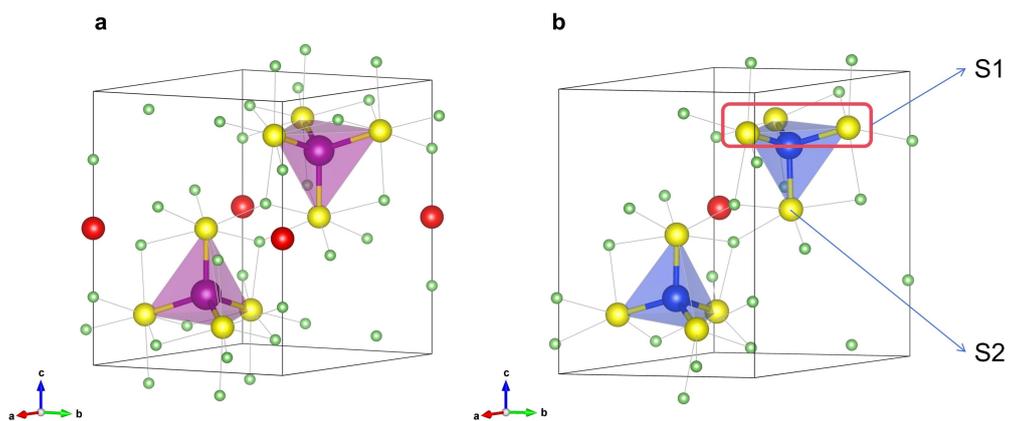

**Fig. S2. The bonding characteristics of (a) LMS and (b) LSS.** The red balls refer to the isolated S atoms. The yellow balls refer to the bonded S atoms. The green balls refer to the Li$^+$ ions. The purple balls refer to the Mn atoms. The blue balls refer to Si atoms.

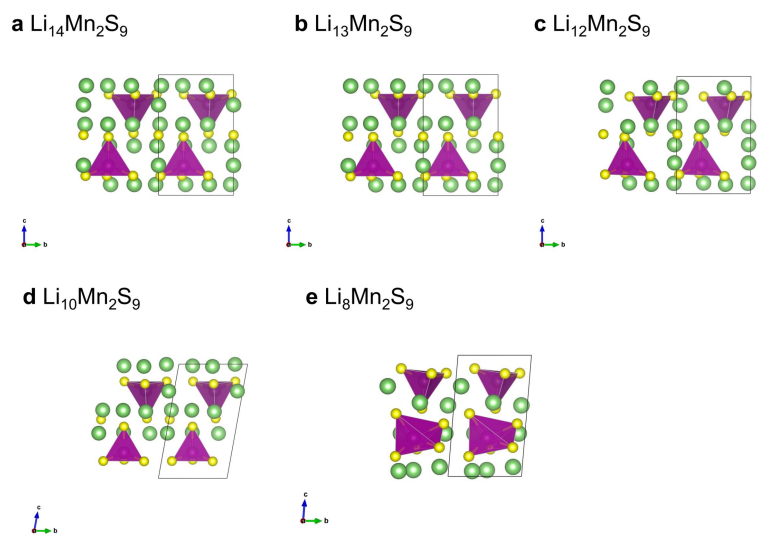

**Fig. S3. The crystal structures for (a) Li$_{14}$Mn$_2$S$_9$, (b) Li$_{13}$Mn$_2$S$_9$, (c) Li$_{12}$Mn$_2$S$_9$, (d) Li$_{10}$Mn$_2$S$_9$, (e) Li$_8$Mn$_2$S$_9$ relaxed by DFT calculations.** The green, purple and yellow balls refer to the Li$^+$ ions, Mn atoms and S atoms.

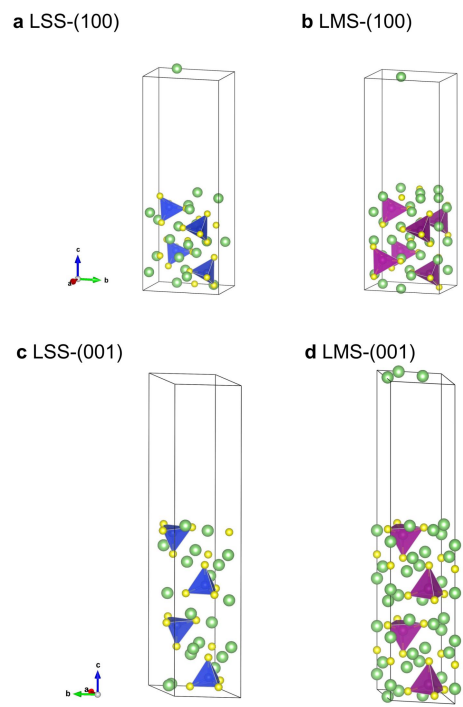

**Fig. S4. The structural models of (a) LSS-(100) surface, (b) LMS-(100) surface, (c) LSS-(001) surface, and (d) LMS-(001) surface.** The yellow, green, purple and blue balls refer to the S atoms, Li$^+$ ions, Mn atoms and Si atoms.

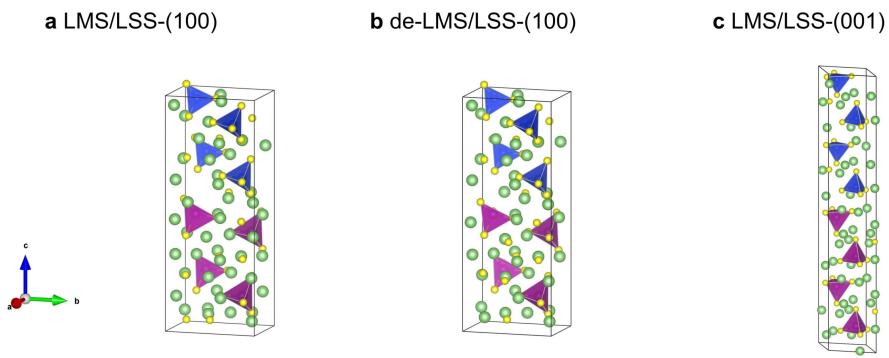

**Fig. S5. The structural models of (a) LMS/LSS-(100), (b) LMS/LSS-(001), (c) de-LMS/LSS-(100).** The yellow, green, purple and blue balls refer to the S atoms, Li⁺ ions, Mn atoms and Si atoms.

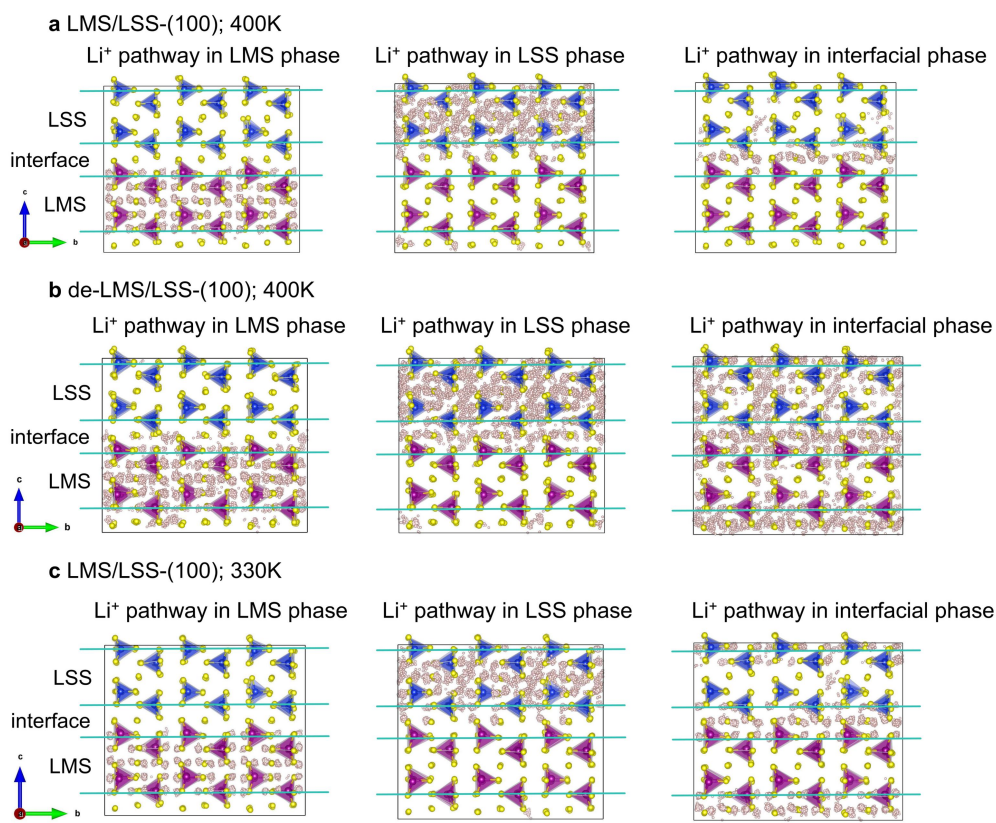

**Fig. S6. The Li⁺ pathways for (a) LMS/LSS-(100) at 400K, (b) de-LMS/LSS-(100) at 400K, (c) LMS/LSS-(100) at 330K.** The yellow, green, purple and blue balls refer to the S atoms, Li$^+$ ions, Mn atoms and Si atoms. The pink balls refer to the Li$^+$ diffusion pathways.

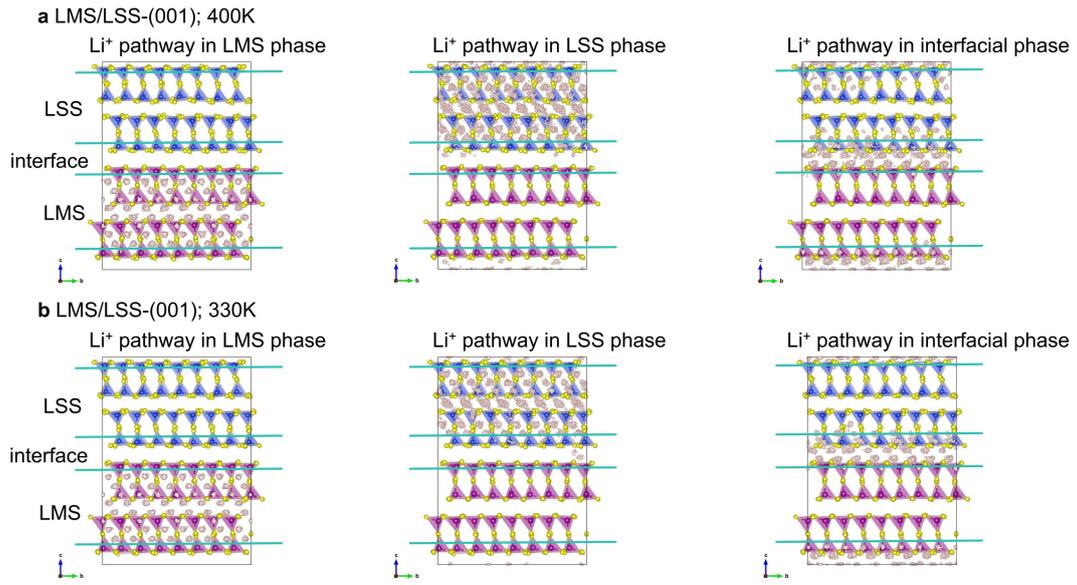

**Fig. S7. The Li⁺ pathways for (a) LMS/LSS-(001) at 400K, (b) LMS/LSS-(001) at 330K.** The yellow, green, purple and blue balls refer to the S atoms, Li$^+$ ions, Mn atoms and Si atoms. The pink balls refer to the Li$^+$ diffusion pathways.

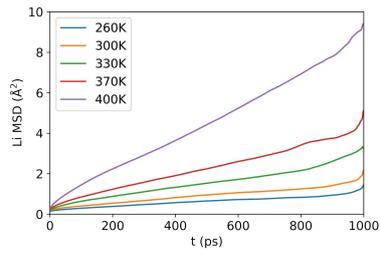

**Fig. S8. The Li⁺ MSDs at different temperatures for the original (100) interface structure.**

**Table S1. Lattice parameters of LMS and LSS.**

| Structure | a (Å) | b (Å) | c (Å) | α (°) | β (°) | γ (°) |
|---|---|---|---|---|---|---|
| $Li_{14}Mn_2S_9$ | 7.04 | 7.04 | 9.87 | 90.00 | 90.00 | 120.00 |
| $Li_{10}Si_2S_9$ | 6.98 | 6.98 | 9.83 | 88.04 | 89.82 | 120.35 |

**Table S2. Total energy of interface structures before and after Mn/Si inter-diffusion (Exchange the position of a Mn atom and a Si atom at the interfaces).**

| Structure | Energy increase value after Mn-Si exchange (eV/cell) |
|---|---|
| LMS/LSS-(100) | 0.525 |
| LMS/LSS-(001) | 0.853 |
| de-LMS/LSS-(001) | -0.048 |

**Table S3. Total number of atoms in unit-cells and supercells.**

| Structure | Total Number of Atoms in unit-cells | Total Number of Atoms in Supercell |
|---|---|---|
| $Li_{14}Mn_2S_9$ | 25 | 75 |
| $Li_{10}Si_2S_9$ | 21 | 84 |
| $Li_{13}Mn_2S_9$ | 24 | - |
| $Li_{12}Mn_2S_9$ | 23 | - |
| $Li_{10}Mn_2S_9$ | 21 | - |
| $Li_8Mn_2S_9$ | 19 | - |
| $Li_6Mn_2S_9$ | 17 | - |
| $Li_4Mn_2S_9$ | 15 | - |
| LMS-(100) | 50 | - |
| LMS-(001) | 50 | - |
| LSS-(100) | 42 | - |
| LSS-(001) | 42 | - |
| LMS/LSS-(100) for MLIP-MD | 92 | 1104 |
| LMS/LSS-(001) for MLIP-MD | 88 | 1320 |
| de-LMS/LSS-(100) for MLIP-MD | 92 | 1472 |

**Table S4. The evaluation results for MLIP models of different interfaces.**

| Model | Force rmse (eV/Å) | Energy rmse (eV/atom) |
|---|---|---|
| LMS/LSS-(100) | 0.087 | 0.0006 |
| LMS/LSS-(001) | 0.099 | 0.0010 |
| de-LMS/LSS-(100) | 0.094 | 0.0008 |

**Note S1. The relevant information on the construction of surface and interface structure models.**

We chose (100) and (001) directions' surface and interface structural models (LMS-(100), LMS-(001), LSS-(100), LSS-(001)) of LMS and LSS for analysis, because the structural frameworks of LMS and LSS exhibit a layered distribution along the c-axis with the surfaces of (100) and (001) directions distributed along and perpendicular to the c-axis respectively, and selecting these two surfaces connected interfaces for simulation enables LMS and LSS to share the same framework structure, reducing the resistance of $Li^+$ ion transport at the interface and is beneficial for directly observing the lithium ion migration behavior within and between layers. Due to the limitations of AIMD computation time and cost, a two-fold unit-cell structure was selected, and a 15 Å vacuum layer in the c-axis direction was chosen for the surface models to ensure that the surfaces do not affect each other. The (100) and (001) interface models (LMS/LSS-(100), LMS/LSS-(001)) are respectively composed of the LMS and LSS surface models in the corresponding direction, which are spliced together. Additionally, the delithiated interface structure for (100) interface (de-LMS/LSS-(100)) is obtained by randomly generating two Li vacancies of LMS within LMS/LSS-(100) model. Within these interface structural models, to determine the distance between LMS and LSS surface models, optimization was performed on configurations featuring varying distances between the LMS and LSS phases and the configurations with the lowest energy were selected.

**Note S2. The information about MLIP model training.**

The MLIP models use the training sets of AIMD simulations above 10 ps at 500K for each unit-cells of interface model, employing parameters derived from the examples provided by the NequIP framework. Three MLIP models of LMS/LSS-(100), LMS/LSS-(001), and de-LMS/LSS-(100) were trained respectively due to different structure and ion conducting characteristics at these three interfaces. The evaluation results of the trained MLIP models were listed in Table S4 confirming their reliability.